\def\beq{\begin{equation}}
\def\eeq{\end{equation}}
\def\bea{\begin{eqnarray}}
\def\eea{\end{eqnarray}}
\def\bq{\begin{quote}}
\def\eq{\end{quote}}
\def    \missEt      {\ifmmode{/\mkern-11mu E_T}\else{${/\mkern-11mu E_T}$}\fi}
\def\BReq {\ifmmode { {\cal B}_{eq}} \else {${\cal B}_{eq}$}\fi}
\def\BRR  {\ifmmode { {\cal B}_{R}} \else {${\cal B}_{R}$}\fi}

\def \lsim{\mathrel{\vcenter
     {\hbox{$<$}\nointerlineskip\hbox{$\sim$}}}}
\def \gsim{\mathrel{\vcenter
     {\hbox{$>$}\nointerlineskip\hbox{$\sim$}}}}

\def\gappeq{\mathrel{\rlap {\raise.5ex\hbox{$>$}}
{\lower.5ex\hbox{$\sim$}}}}

\def\lappeq{\mathrel{\rlap{\raise.5ex\hbox{$<$}}
{\lower.5ex\hbox{$\sim$}}}}

\def\bbz{fa Z \kern-8.9pt Z}
\documentstyle [12pt,epsfig]{article}
  \newcommand{\ccaption}[2]{
    \begin{center}
    \parbox{0.85\textwidth}{
      \caption[#1]{\small{\it{#2}}}
      }
    \end{center}
    }
\begin{document}
\thispagestyle{empty}
\vspace*{-1cm}
\begin{flushright}
{CERN-TH/97-101} \\
{\tt hep-ph/9705287} \\
\end{flushright}
\vspace{1cm}
\begin{center}
{\large {\bf \sc What if Charged Current Events at Large Q$^2$ \\ are Observed
at HERA?}} \\                                                 
\vspace{.2cm}                                                  
\end{center}

\begin{center}
\vspace{.3cm}
{\bf G. Altarelli\footnote{Also at Universit\`a di Roma III, Rome, Italy.}, 
G.F. Giudice\footnote{On leave of absence from INFN, Sez. di Padova, Italy.} }
 and                                                     
{\bf M.L. Mangano}\footnote{On leave of absence from INFN, 
                            Sez. di Pisa, Italy.} 
\\                          
\vspace{.5cm}
{ Theory Division, CERN, CH-1211, Gen\`eve 23, Switzerland} \\
\end{center}                              

\vspace{2cm}

\begin{abstract}
An excess of events at large $Q^2$ with a positron in the final state has
been observed at HERA which, if confirmed, would be a signal of new
physics. It is not clear at present if a signal of comparable rate is also
seen in the charged current channel (with an antineutrino in the final
state). In this note we analyse the implications of the presence
of such a signal in models of new physics based on contact terms,
leptoquarks and squarks with R-violating decays. We find that in all
cases the most likely possibility is that the  charged current
signal is absent. As a
consequence if this signal is present the resulting indications are very
selective. In particular for squarks only charged current events with
multi-quark final states
are possible with quite definite predictions on the spectrum of supersymmetric
particles.
 \end{abstract}
\vfill 
\begin{flushleft}
CERN-TH/97-101\\
May 1997\\    
\end{flushleft}

\newpage

\section{Introduction}
As well known by now, both HERA experiments, H1~\cite{h1} and ZEUS~\cite{zeus}, 
have reported an excess
of events with respect to the Standard Model (SM) expectation in
$e^+p\rightarrow e^+X$ at very large and until now unexplored values of
$Q^2\gappeq 1.5 \times
10^4$ GeV$^2$. The limited statistics and the somewhat imperfect matching of
the
H1
and ZEUS findings still leave ample margins of doubt on the reality of
this new
physics indication. However possible explanations invoking either
additional
contact terms in the effective Lagrangian or a leptoquark of mass
$M\sim200$ GeV,
presumably an up-type squark ($\tilde q$) decaying with R-parity violation,
have been studied in
detail~\cite{Altarelli97,HERARPV,HERAlq,HERAcomp1,HERAcomp2,HERAcomp3}. 
At present it is still not clear if a similar signal is also present in   
the charged-current (CC) channel, {\it i.e.} with an antineutrino in the final
state. Only the
H1 experiment reports 4 possible events in this channel at $Q^2\gappeq 
1.5 \times
10^4$ GeV$^2$ with
$1.77\pm0.87$ expected, too few to
reach
any definite
conclusion, while the ZEUS analysis of the CC channel is still in
progress.
The
presence or absence of a simultaneous CC signal is extremely significant
for the
identification of the underlying new physics (as it would also be the case
for
the result of a comparable run with an $e^-$ beam, which however is further
away in time). The HERA run presently under way, still with an $e^+$ beam, will
soon tell us more on the reality of the new physics evidence and on whether or
not a signal is also present in the CC channel. In view of this, in this note
we consider in detail the implications for the CC channel of the various
proposed solutions of the HERA effect. We will confirm that in most of the
cases the CC signal is not expected to arise. But if it is present at a
comparable rate as for the NC signal, the corresponding indications are very
selective. A contact term would be practically  excluded. If the associated jet
is as sharp as in the neutral current (NC) case,
suggesting one single final-state quark, then this would point towards a
leptoquark of peculiar structure, either with couplings explicitly violating
the electroweak $SU(2)\bigotimes U(1)$ symmetry, or perhaps with couplings to an
antineutrino and a charm quark. On the other hand a squark of the charm
($\tilde c$) or top ($\tilde t$)
type could lead to a CC signal with multijets in the final state arising from
the cascade into quark plus gaugino, the latter decaying via R-violating
couplings to a neutrino
plus two quarks. Well specified small regions of the parameter space should be
selected in order to produce a CC signal with the required features. The    
relevant distributions for this type of decay will be studied in some detail.
In the following we will first study the case of contact terms, then we will 
discuss leptoquarks in general and finally we will deal with the squarks with
R-violating interactions. 
                          
\section{Contact Terms}
From the present NC data one cannot exclude a non resonant solution, although
the                 
H1
data do indeed favour a leptoquark resonance. Thus at the moment one
can     
obtain
a reasonable, although not very good, fit of the available distributions in
the NC
channel in terms of vector contact terms of the general form
\beq
\Delta L=\frac{4\pi\eta_{ij}}{(\Lambda^\eta_{ij})^2} \; \bar
e_i\gamma^{\mu}e_i \; \bar q_j\gamma_{\mu}q_j     
\label{1}             
\eeq with $i,j=L,R$ and $\eta$ a $\pm$ sign. Strong limits on these
contact
terms
are provided by LEP~\cite{Opal96,Opal97}, Tevatron~\cite{Bodek96} and 
atomic parity violation (APV) experiments~\cite{wood,mas}. But
for                                               
example a parity conserving combination
$({\bar e}_L\gamma^\mu e_L)(  {\bar u}_R\gamma_\mu u_R) +
( {\bar e}_R\gamma^\mu e_R )( {\bar u}_L\gamma_\mu u_L)$
with
$\Lambda^+_{LR}=\Lambda^+_{RL}\sim 3$ TeV leads to an acceptable fit to
the HERA
data and is compatible with the existing
limits~\cite{Altarelli97,HERAcomp1,HERAcomp2}. 
For this contact term,                                                    
even in
the limit of SU(2) symmetry, there is no need for an associated CC signal,
because
the current ${\bar e}_L\gamma^\mu e_L$  
could be part of a singlet ${\bar e}_L\gamma^\mu e_L +
{\bar \nu}_L\gamma^\mu \nu_L $.
We
note
that for contact terms it is natural to assume the validity of the
$SU(2)\bigotimes U(1)$ symmetry, because they are associated with physics
at a
large energy scale. 

Suppose now that there is a CC signal of comparable rate as the NC one. In
the
$SU(2)\bigotimes U(1)$ limit, restricting us to family diagonal quark
currents in order to minimise problems with the occurrence of flavour
changing neutral currents, the only possible vector contact term with
valence quarks (and no Cabibbo suppression) is of the form 
\beq
\Delta L_{CC}=\frac{4\pi\eta}{\Lambda^2_{\eta}}
{\bar e}_L\gamma^\mu \nu_L {\bar u}_L\gamma_\mu d^\prime_L + {\rm h.c.}
\label{due}
\eeq {\it i.e.} the product of two isovector currents. 
Here $d_L^\prime$ is the left-handed d-quark current eigenstate, related
to the mass eigenstate by the Cabibbo-Kobayashi-Maskawa (CKM) matrix.

It is simple to see that such
terms cannot have a sufficent magnitude.
On the one hand the rate for events with antineutrinos in the final state
would be modified by the presence of the contact term of the form in
eq.~(\ref{due})
by an amount that can easily be estimated because the chiral structure of
the additional term is the same as in the SM. Neglecting the
charm-quark contribution we have
\beq
\frac{d\sigma}{dQ^2}(\Lambda)=\frac{d\sigma_{SM}}{dQ^2} \left[
1- \frac{2\eta \sin^2\theta_W}{\alpha}
\frac{(Q^2+M^2_W)}{\Lambda^2}\right]^2~.
\label{tre}
\eeq
Inserting the values of the electromagnetic coupling $\alpha$, the weak
mixing angle $\theta_W$ and the W mass, for $Q^2=15000~GeV^2$ we find
\beq
\frac{d\sigma}{dQ^2}(\Lambda)=\frac{d\sigma_{SM}}{dQ^2} 
\left[ 1 -\eta ~(1.1~{\rm TeV}/\Lambda)^2\right]^2  \label{quattro}
\eeq           
Thus values of $\Lambda$ in the range between 1 and 1.5 TeV and $\eta=-1$
would be required to produce a sizeable excess.
     
On the other hand, the
scale
$\Lambda$ associated with this operator is strongly constrained by at
least two
experimental facts: lepton-hadron universality of weak charged currents
and electron-muon universality in charged-pion decays. Consider the 
strength of
the
four fermion interaction responsible for muon decay and compare it to that
of the                                   
similar term with the muon current replaced by the $u\rightarrow d'$
current. The
only room for a new interaction as in eq.~(\ref{due}) is within the allowed
discrepancy
from
unitarity of the CKM matrix. From the experimental values~\cite{pdg}:
\beq                                                      
|V_{ud}|=0.9736\pm0.0010,~~~|V_{us}|=0.2205\pm0.0018,
~~~~|V_{ub}|=0.0033\pm0.0009
\label{3}
\eeq one finds 
\beq |V_{ud}|^2+|V_{us}|^2+|V_{ub}|^2=0.9965\pm0.0022 \label{4}
\eeq As a consequence, at 1.64$\sigma$, one obtains the constraint
\beq 0.9929<1-\frac{2\sqrt{2}\pi \eta }{G_F \Lambda_{\eta}^2}<1.0001 \label{5}
\eeq Thus at 90\% CL one finds that $\Lambda_+>10$ TeV and
$\Lambda_->87$ TeV.

We now consider $e-\mu$ universality in charged pion decay. We assume that
there
is a contact term for electron but not for muon currents. Then the ratio
$R=\Gamma
(\pi^{-} \rightarrow e \bar
\nu)/\Gamma (\pi^{-} \rightarrow \mu \bar \nu)$ would deviate from its
SM value by
\beq R=R_{SM} (1-\frac{2\sqrt{2}\pi \eta }{G_F \Lambda_{\eta}^2})
\label{6}
\eeq 
According to ref.~\cite{mar}
the present experimental value of $R_{exp}/R_{SM}$ is
given by:
\beq R_{exp}/R_{SM}=0.9966\pm0.0030 \label{7} \; .
\eeq 
As a consequence, at 1.64 $\sigma$ we find   
\beq 0.9917<1-\frac{2\sqrt{2}\pi \eta }{G_F \Lambda_{\eta}^2}<1.0015 \; .
\label{8}                                                                
\eeq 
This leads to $\Lambda_+> 10$ TeV and $\Lambda_- >
23$ TeV at 90\% CL.                               

We conclude that vector contact terms of this type could at most lead to a
CC signal below
the
percent level with respect to the NC one. This statement
remains
true
even if we relax the $SU(2)\bigotimes U(1)$ symmetry. For example, if we
impose
that
the axial hadronic current vanishes, in order to evade the charged-pion
constraint,
the universality bound is still valid for the vector hadronic current with
respect
to the muon vector current. Similarly, trying to restore $e-\mu$
universality by also allowing a corresponding contact term with the
electron current replaced by a muon current with exactly the same
coupling would also fail. First of all there are strong limits from the
precisely measured ratio $R_\nu$ of NC to CC rates in $\nu_{\mu}$ induced deep
inelastic reactions~\cite{pdg}. Secondly, the limit from the comparison with
muon decay remains valid. Moreover we cannot also introduce a contact term
which contributes to muon decay, because the determination of $G_F$ would be
affected and precision tests of the SM practically exclude that. Furthermore,
very stringent limits from $e^+e^-\to \mu^+\mu^-$ have been set by
OPAL~\cite{Opal96}.
                                                                       
The possible scalar or tensor currents arising from an $SU(2)\otimes U(1)$
invariant theory which can contribute to valence-parton CC processes are
\beq
{\cal L}=
\frac{4 \pi}{\Lambda_S^2}({\bar e}_R \nu_L)( {\bar u}_R d_L)+
\frac{4 \pi}{\Lambda_{S^\prime}^2}({\bar e}_R \nu_L)( {\bar u}_L d_R)+
\frac{4 \pi}{\Lambda_T^2}({\bar e}_R\sigma^{\mu \nu} \nu_L)( {\bar u}_R 
\sigma_{\mu \nu}d_L) ~,
\eeq
while the operator $({\bar e}_R\sigma^{\mu \nu} \nu_L)( {\bar u}_L 
\sigma_{\mu \nu}d_R)$ identically vanishes. The scalar interactions are
strongly limited by $e$--$\mu$ universality in pion decays~\cite{sha}, 
\beq
\Lambda_{S,S^\prime} > 500 ~{\rm TeV} ~,
\label{scalar}
\eeq
because 
they do not lead to electron-helicity suppression, in contrast with
the SM case. Even introducing a muon counterpart to the
electron contact term could not help in this case barring an
unbelievable level of
fine tuning. The tensor interaction can be dressed into a scalar
interaction of effective strength~\cite{vol} 
\beq
\frac{1}{\Lambda_{S~eff}^2}\simeq
-\frac{\alpha}{\pi}\log\left( \frac{\Lambda_T^2}{M_W^2}\right) ~  
\frac{1}{\Lambda_T^2}~,
\eeq
with the
exchange of a photon between the electron and the quark fields.
Then lepton universality in pion decays sets the limit         
\beq
\Lambda_T > 90 ~{\rm TeV} ~.
\label{tensor}
\eeq
 
Considering now also  CC processes involving sea quarks, we can
introduce a contact term for second generation quarks
\beq
\Delta L_{CC}=\frac{4\pi\eta}{\Lambda^{(2)2}_{\eta}}(
{\bar e}_L\gamma^\mu \nu_L)( {\bar c}_L\gamma_\mu s^\prime_L) + {\rm h.c.}
\label{charm}
\eeq 
Clearly since the strange sea in the proton is small one needs relatively
small values of $\Lambda$ in order to produce a sufficiently large effect.
A detailed study shows that one needs $\Lambda\sim0.8-1$~TeV with
$\eta=-1$ in order to obtain an increase by a factor of two with respect
to the SM at $Q^2=15000~GeV^2$. Contact terms with $\eta=1$
give negative interference at HERA and one would be 
forced to
take $\Lambda$ very small in order to take advantage of the contact term
squared. However such a term would lead to a disagreement 
with the data at low $Q^2$.

Bounds on the scales $\Lambda_{\eta}^{(2)}$ can be derived
from
lepton universality in $D$ decays~\cite{pdg}
\beq
R^D_{e/\mu}\equiv \frac{\Gamma (D^0\to K^-e^+\nu )}{1.03\times 
\Gamma (D^0\to K^-\mu^+\nu )}=
1.09 \pm 0.09~.
\label{rdemu}
\eeq
Here the factor 1.03 takes into account the phase-space suppression of
the muon channel. The operator in eq.~(\ref{charm}) predicts a deviation
from universality
\beq  \label{eq:Denu}
R^D_{e/\mu}= 1-\frac{2\sqrt{2}\pi \eta }{G_F \Lambda^{(2)2}_{\eta}}~,
\eeq
and it is therefore constrained by eq.~(\ref{rdemu}) to satisfy
\beq
\Lambda_+^{(2)}> 3.6 ~{\rm TeV} ~~{\rm at~90\% ~CL},
\eeq
\beq
\Lambda_-^{(2)}> 1.8 ~{\rm TeV} ~~{\rm at~90\% ~CL}.
\eeq

Bounds that do not rely on $e-\mu$ universality can
be obtained from the unitarity of the CKM matrix. One can compare
$|V_{cs}|$
obtained from unitarity by $|V_{cs}|=\sqrt{1-|V_{us}|^2-|V_{ts}|^2}$
with the value directly measured from $D\rightarrow K e \nu$, which is
affected by the
contact term 
in eq.~(\ref{charm}). Using $|V_{ts}|=0.040\pm0.004$ and
$|V_{us}|=0.2205\pm0.0018$~\cite{pdg}, we have                   
$|V_{cs}|=\sqrt{1-|V_{us}|^2-|V_{ts}|^2}=0.9746\pm0.0004$. This is to
be
compared with $|V_{cs}|=1.01\pm0.18$ from $D\rightarrow K e \nu$~\cite{pdg}. 
As a result we find
\beq
\Lambda_+^{(2)}> 1.2~{\rm TeV} ~~{\rm at~90\% ~CL},
\eeq
\beq
\Lambda_-^{(2)}> 1.1 ~{\rm TeV} ~~{\rm at~90\% ~CL}.
\eeq
If we introduced an $e-\mu$ symmetric combination of contact terms a
comparison between the value of $|V_{cd}|$~\cite{pdg} extracted from charm
production off $d$-valence quarks in (muon) neutrino-scattering
experiments and                                                    
the value of $|V_{cd}|$ extracted from unitarity plus knowledge of
$|V_{ud}|$ and $|V_{td}|$~\cite{pdg} would set on the common coupling the
limits 
\beq                                 
\Lambda_+^{(2)}> 1.7~{\rm TeV} ~~{\rm at~90\% ~CL},
\eeq
\beq
\Lambda_-^{(2)}> 1.9 ~{\rm TeV} ~~{\rm at~90\% ~CL}.
\eeq
These limits are at the level of the present sensitivity of LEP2 
experiments~\cite{Opal97},        
but should be improved after the next run.

Scalar contact terms involving second-generation quarks are constrained
by leptonic decays of $K^+$, $K^0$, and $D^0$ and cannot give a significant
contribution to CC events at HERA.

In conclusion it appears very difficult 
to accomodate a CC signal at HERA in the framework of contact terms. 

\section{Leptoquarks}
Let us now consider a scalar leptoquark resonance that is coupled both to
$e^+d$ and to $\bar \nu u$ so that it can generate both NC and CC events from
valence (note that $e^+u$ has charge +5/3 and cannot go into $\bar \nu q$). A
vector leptoquark has a much larger cross section at the Tevatron than a scalar
leptoquark~\cite{Bluemlein96}, and current~\cite{D0LQ} or upcoming limits 
should be able to rule out this possibility.  Assuming that the symmetry under 
$SU(2)\bigotimes U(1)$ is conserved, the virtual leptoquark exchange
gives a CC contribution to the low-energy effective Lagrangian of the form
\beq
{\cal L}=\frac{\lambda_u \lambda_d}{M^2}(\bar e_R d_L)(\bar u_R \nu_L)
+{\it h.c.}
\label{9}
\eeq
Here $\lambda_u$ and $\lambda_d$ are the (real) couplings of a leptoquark with
mass $M$ to the $\bar \nu_L u_R $ and $\bar e_R d_L$ currents,
respectively.
This interaction corresponds to the transition $e^+_L d_L \rightarrow \bar
\nu_R
u_R$
which has $T=-1/2$ both in the initial and final states. 
At low energies, the leptoquark exchange induces a contribution to
$\pi \to e \bar \nu$ which is not helicity suppressed. After Fierz
rearranging eq.~(\ref{9}), we can translate the limit in eq.~(\ref{scalar})
into an upper bound on the leptoquark branching ratio into neutrinos,
$BR(LQ\to \bar \nu u)\simeq \frac{\lambda_u^2}{\lambda_d^2}<5\times
10^{-6}$ for $M\sim 200$ GeV and $\lambda_d\sim 0.04$~\cite{Altarelli97}.
This clearly excludes any observable CC signal, if the scalar leptoquark
produced from valence quarks has gauge-invariant couplings.

An alternative is to break $SU(2)\bigotimes U(1)$ and assume that the
leptoquark exchange induces an effective interaction of the form
\beq
{\cal L}=\frac{\lambda_u \lambda_d}{M^2}(\bar e_L d_R)(\bar u_R \nu_L)
+{\it h.c.}
\label{10}
\eeq
Note that in the transition $e^+_R d_R \rightarrow \bar \nu_R u_R$ the
initial
state has $T=+1/2$ while the final state has $T=-1/2$. In this case the low
energy
effective interaction gives a contribution to $\pi \rightarrow e \bar \nu$
which
is helicity suppressed and so could be acceptable. In fact by Fierz
rearrangement
we have
\beq
{\cal L}=-\frac{\lambda_u \lambda_d}{2~M^2}(\bar e_L \gamma^{\mu} \bar
\nu_L)(\bar
u_R
\gamma_{\mu} d_R )+{\it h.c.}~, 
\eeq
where the minus sign comes from the anticommutative properties of the fermionic
fields.                             
The limits discussed in eqs.~(\ref{6})--(\ref{8}) give
\beq
|\lambda_u \lambda_d |< 1\times 10^{-2}\left( \frac{M}{200~{\rm GeV}}\right)^2
~~~~{\rm if}~\lambda_u\lambda_d>0~,
\label{lqpos}
\eeq
\beq
|\lambda_u \lambda_d |< 2\times 10^{-3}\left( \frac{M}{200~{\rm GeV}}\right)^2
~~~~{\rm if}~\lambda_u\lambda_d<0~.
\label{lqneg}
\eeq
We recall that the observed rate of the NC events at HERA
requires~\cite{Altarelli97}                          
$|\lambda_d|\simeq 4\times 10^{-2}/\sqrt{1-{\cal B}_{\nu u}}$, where
${\cal B}_{\nu u}\equiv BR(LQ\to \bar \nu u)=(1+\lambda_d^2/\lambda_u^2)^{-1}$. 
Equations~(\ref{lqpos})
and (\ref{lqneg}) allow ${\cal B}_{\nu u}$ values as large as 73\% if
$\lambda_u$ and $\lambda_d$ have the same sign and up to 38\% if they have
opposite sign.

It is interesting to speculate on how the leptoquark couplings could violate
gauge invariance. Since $SU(2)\otimes U(1)$ is broken only by the Higgs vacuum
expectation value (VEV), we have to assume that the leptoquark couples to the
quark-lepton current through some higher-dimensional operator. Indeed, the
dimension-five interactions
\beq
{\cal L}=\frac{\hat \lambda_u}{\cal M}\Phi \bar u_R H^T(-i\sigma_2)\ell_L
+\frac{\hat \lambda_d}{\cal M}\Phi \bar d_R H^\dagger \ell_L +{\rm h.c.}
\label{efflag}
\eeq
give rise to the desired couplings
\beq
\lambda_u=\hat \lambda_u \frac{v}{\cal M},~~~
\lambda_d=\hat \lambda_d \frac{v}{\cal M},~~~
\eeq
after the Higgs field acquires its VEV, $\langle H\rangle =(0,v)^T$. 
Equation~(\ref{efflag}) determines the $SU(3)\otimes SU(2)\otimes U(1)$ quantum 
numbers of the scalar leptoquark field $\Phi$ to be 
the same as those of the right-handed
up quark.

Appropriate values of $\lambda_u$ and $\lambda_d$ are achieved when the
mass scale ${\cal M}$ of the effective interaction is below about 10 TeV.
This scale is low enough to require some further discussion about the 
underlying dynamics. In the context of perturbative physics, it is possible
to generate the effective Lagrangian in eq.~(\ref{efflag}) via exchange
of a single non-chiral fermion $X$ with the same quantum numbers of the
left-handed quark doublet, and with the following interactions:
\beq
{\cal L}=\left[ y_\ell\Phi \bar X_R \ell_L   +
y_u \bar u_R H^T(-i\sigma_2)X_L +   y_d \bar d_R H^\dagger X_L
+{\rm h.c.}\right] -{\cal M} \bar XX~.
\label{lag}
\eeq
After integrating out the heavy field $X$, we recover eq.~(\ref{efflag}) with
$\hat \lambda_{u,d}=y_\ell ~y_{u,d}$. We have no satisfactory
explanation for why the
field $X$ couples only to first generation quarks and leptons. This model
is just meant to give an illustrative example of a possible leptoquark
which can produce CC events at HERA.

The interactions in eq.~(\ref{lag}) generate, after electroweak symmetry
breaking, a mass mixing between ordinary
quarks and the field $X$. Keeping just the leading effects in $1/{\cal M}$,
the mixing is only among right-handed particles and it induces a coupling
of the $W$ boson with the first-generation hadronic right-handed current,
suppressed by a factor $\rho\equiv (\lambda_u\lambda_d)/y_e^2$ with respect 
to the usual left-handed current. Direct limits on hadronic right-handed
currents from deep-inelastic scattering experiments give $\rho^2<0.009$
at 90\% CL~\cite{abr}. Neutral currents coupled to the $Z$ boson are also
modified in their isospin
part $J_3^\mu$, but of course not in their electromagnetic part. The vector
and axial-vector couplings of a first generation quark $f$ become
\beq
v_f =T_f\left( 1+\frac{\lambda_f^2}{y_e^2}\right) -2Q_f\sin^2\theta_W~,~~~
a_f =T_f\left( 1-\frac{\lambda_f^2}{y_e^2}\right) ~,
\label{vaf}
\eeq
where $T_f$ and $Q_f$ are the third-component isospin and electric charge
of the fermion $f$. APV experiments constrain the
new contributions in eq.~(\ref{vaf}). Comparing the measured and predicted
values of the cesium ``weak charge", $Q_W^{exp}=-72.11\pm 0.93$~\cite{wood}, 
$Q_W^{SM}=-73.2\pm 0.2$~\cite{mas}, we obtain an allowed range for
$\lambda_u$ and $\lambda_d$ given at 90\% CL by
\beq
-1.3\times 10^{-2}<\left( \frac{\lambda_d}{y_e}\right)^2 -0.89
\left( \frac{\lambda_u}{y_e}\right)^2 <2.2\times 10^{-3}~.
\label{cesio}
\eeq
Using $|\lambda_d|\simeq 4\times 10^{-2}/\sqrt{1-{\cal B}_{\nu u}}$, 
we find that
eq.~(\ref{cesio}) allows values of ${\cal B}_{\nu u}$
as large as 77\%,                            
for  $y_e\sim1$. Precision measurements at LEP do not constrain the model
further, and therefore the possibility of CC events at HERA from this
particular leptoquark
with non-gauge-invariant effective couplings is still allowed.

Another viable alternative is a leptoquark which couples simultaneously to
the $\bar e_R^{(1)} q_L^{(1)}$ and $\bar \ell_L^{(i)} u_R^{(2)} $ currents
($i=1,2,3$).
Here we have specified the generation indices of the different fields.
If CC events were observed at HERA and such a leptoquark was responsible
for them, we expect the striking signature of leptonic $D$ decays with
rates much larger than in the SM
if $SU(2)\bigotimes U(1)$ is respected:
\beq
BR(D^0\to e^{\pm(1)} e^{\mp(i)})\simeq 1\times 10^{-6}
\frac{{\cal B}_{\nu c}}{(1-{\cal B}_{\nu c})^3}
\left( \frac{200~{\rm GeV}}{M}\right)^4~~~~i=1,2~.
\eeq
The present experimental limits $BR(D^0\to e^{\pm} e^{\mp})<1.3\times 10^{-5}$
and $BR(D^0\to e^{\pm} \mu^{\mp})<1.9\times 10^{-5}$~\cite{pdg} allow  large
values of the leptoquark branching ratio into ${\bar \nu} c$, ${\cal B}_{\nu 
c}$.
The leptoquark under consideration belongs to an $SU(2)$ doublet and the
mass of its partner is constrained by the electroweak $\rho$ parameter.
Allowing for a new physics contribution $\Delta \rho < 1\times 10^{-3}$,
we find that the second leptoquark must be lighter than 250 GeV. This
state has electric charge 5/3, gives rise only to NC events and it is produced
by positron scattering off up quarks. Its cross section is, in the worst
case (for $M=250$ GeV), only 4 or 5 times smaller than the cross section
of its weak partner, allegedly produced at HERA.
                                                                         
Let us consider 
the production of a leptoquark resonance from sea quarks.
Production from $\bar u$ or $\bar d$ quarks in the sea is
excluded~\cite{Altarelli97,HERAcomp1} by the $e^- p$ HERA data.
In the case of a leptoquark produced
in the $e^+ s$ channel, the existence of
a ${\bar \nu} u$ final state is severely constrained by 
$e-\mu$ universality in $K \rightarrow \ell \nu$~\cite{david}.
On the other hand,
the possibility of a ${\bar \nu} c$ final state is still allowed. 
This leads to a remarkable signature in leptonic $D_s$ decays
\beq            
BR(D_s^-\to e^-\bar \nu)\simeq 6\times 10^{-3}\frac{{\cal B}_{\nu 
c}}{(1-{\cal B}_{\nu c})^3}
\left( \frac{200~{\rm GeV}}{M}\right)^4~~~~i=1,2~.                         
\eeq
We are not aware of any existing experimental limit on this quantity.

To conclude, we recall that
a leptoquark with branching ratio equal to 1 in
$e^+q$ is practically excluded by
the Tevatron, 
as discussed in more detail in the next section. 
Therefore on one hand some branching fraction in the CC channel
is needed. On the other hand, we find that there is limited space
for the possibility that a leptoquark can generate a CC signal at HERA with one
single parton quark in the final state. This occurrence would indicate
$SU(2)\bigotimes U(1)$ violating couplings and corresponding higher dimension
effective operators with Higgs fields, or couplings to a current containing the
charm quark.

\section{Squarks with R-Parity Violating Decays}
\subsection{General Constraints}
Perhaps the most attractive possibility is that the HERA signal is a
manifestation of supersymmetry~\cite{MSSM}, 
in the specific form of a squark with an
R-violating coupling~\cite{RPth}. 
Production of squarks at HERA via $R$-violating interactions has been an area
of active study for quite some time (see, e.g., 
refs.~\cite{Hewett,Kon,Dreiner}.)
Recently this possibility has been reconsidered in        
detail~\cite{Altarelli97,HERARPV,HERAcomp1} in the light of the new HERA
findings. The
R-violating coupling is of the form $\lambda^{\prime}_{1jk} L_1 Q_j D^c_k$ and,
given the present experimental limits on such interactions, the possible
production channels are $e^+d \rightarrow \tilde c_L$, $e^+d \rightarrow \tilde
t_L$, and with more marginal chances, 
$e^+s \rightarrow \tilde t_L$~\cite{Altarelli97}. 
Such squarks are very       
particular leptoquarks and, in fact, their decay into the final state
$\bar \nu u$ is not allowed. However, it has been shown that in all cases
R-conserving decay channels could have competing
branching ratios with the R-violating ones~\cite{Altarelli97}. 
This fact not only makes the squark option more easily   
compatible with the Tevatron bounds, but also offers the possibility of inducing
a CC signal arising from some special R-conserving decay channels. 

Before starting the discussion of the possible decay chains,
we first introduce some notation and discuss current constraints on these
scenarios. We denote by \BReq\ the branching ratio
for the R-violating $e^+q$ decay mode, and $\BRR=1-\BReq$ the
branching ratio into the R-conserving ones. We also define                     
$\lambda^{q}_0=\lambda^{\prime}_{1iq} \cdot \sqrt{\BReq}$ 
to be the variable whose strength is directly      
measured by the NC event rate. We recall that fits to the H1 and ZEUS data 
give~\cite{Altarelli97} $\lambda^{d}_0=0.04$ (prodution on $d$ quarks), and
$\lambda^{s}_0=0.3$  (production on $s$ quarks). These estimates are
clearly affected by the statistical uncertainty on the observed signal, as well
as by uncertainties in the theoretical calculation of the production cross
section. 
Recent calculations of NLO corrections to the leptoquark cross section at
HERA~\cite{Kunszt97} show a $K$-factor correction of the order of 30\%
relative to the Born evaluation used in ref.~\cite{Altarelli97}, and a residual
uncertainty due to scale variations of the order of $\pm 10\%$. 
Accounting for these effects and for the statistical uncertainty of the
observed signal, we shall consider values of $\lambda^{d}_0$
($\lambda^{s}_0$) in the range $0.03-0.04$ ($0.2-0.3$).
                                                              
It is important to notice that, even in absence of constraints from CC events,
improved data from the Tevatron~\cite{D0LQ} on one side and from 
APV~\cite{wood}                       
on the other considerably reduce the window for the
explanation based on $e^+d \to \tilde c$ or $\tilde t$ (and, more in general,
for all leptoquarks). Consistency with the Tevatron demands a value of \BReq\
smaller than 1. 
In fact, the most recent NLO estimates of the squark and leptoquark
production cross sections~\cite{Spira96,Spira97} allow to estimate that at
200~GeV approximately 6--7 events with $e^+e^-jj$ final states
should be present in the combined CDF and D0 data
sets. The absence of event candidates should allow to exclude a value of 
${\cal B}_{eq}^2\gsim 0.5$ at 95\%CL. This constraint could even be tighter,
considering that 
a gluino with mass not much larger than the squark mass would further increase
the squark production rate.
To be conservative, we shall in any case demand $\BReq\lsim 0.75$.
As for the lower limit on \BReq,                                  
the new experimental result on APV in cesium quoted in the previous section and
the relation 
\beq         
\Delta C_{1d}=+\frac{(\lambda^{\prime}_{1id})^2\sqrt{2}}{8m_{\tilde q_i}^2 G_F}
\eeq                                      
(with $C_{1d}$ defined in
ref.~\cite{pdg}, p.87), lead to the bound 
\beq
\vert \lambda^{\prime}_{1id}\vert<0.055\left(\frac{m_{\tilde q_i}}{200~{\rm GeV}}
\right)
\eeq
 at 90\%CL. In turn, using the relation
 $\vert\lambda^{\prime}_{1id}\vert=(0.03-0.04)/\sqrt{\BReq}$,
 this translates into $\BReq\gsim 0.3-0.5$.
Thus, on rather general grounds only the narrow window $0.3\sim 0.5 \lsim \BReq
\lsim 0.75$ is left.

We come now to the study of the decay processes. We recall first
that the required balance of R-violating and R-conserving decay channels can be
obtained if we assume that the mode $\tilde c \rightarrow s \chi^+$ is
forbidden by phase space, that is $m_{\chi^+}>200$ GeV, while the channel
$\tilde c \rightarrow c \chi^0_1$ is allowed and suitable values of the
relevant parameters are selected. In the case of the stop decays,
we observe that the $t\chi^0$ mode is closed, due to the large top mass.
The $b\chi^+$ mode can naturally compete with the R-violating one for a
large range of chargino masses if the production involves an $s$ quark, as in
this case $\lambda^{\prime}$ has a strength of the order of the EW coupling.
For production on a $d$ quark $\lambda^{\prime}$ is very small, 
and the chargino mass needs to be fine tuned 
in order to have a
sufficient phase-space suppression of the otherwise large R-conserving decay
width.                             

In the case of $\tilde c$ production, the most promising decay mode for a CC
signal is the chain\footnote{While our study was performed by considering
decays to all possible $\chi^0_i$ states ($i=1,\dots,4$), it turns out that
only decays to the lightest neutralino are relevant and evade the overall
constraints set by the request of a sizeable CC signal.}:
\beq                                         
\tilde c \to c \chi^0 \to 
        c \nu_e \tilde{\nu}_e \to c \nu_e q \bar q' \; ,
\label{12}                                                        
\eeq
where in the last step the R-violating coupling is involved. 
The $\nu\tilde\nu$ decay of the neutralino competes with the analogous decay
into $\ell\tilde\ell$. 
In order to maximize the CC decay mode, we assume a large mass for all right
components of the charged sleptons. Furthermore, 
we shall assume for simplicity all sneutrino species to
be degenerate, and the standard $SU(2)$ relation between
the masses of the left components of charged sleptons and of 
sneutrinos:
\beq       \label{eq:massrel}
  m_{\tilde{\ell}_L}^2=m_{\tilde{\nu}_{\ell}}^2+\vert \cos 2\beta\vert m_W^2 \;.
\eeq                                                                     
The relative branching ratios into the neutral and charged slepton channels 
are given by:
\beq
   \frac{BR(\chi^0_i\to \ell \tilde{\ell}_L)}
        {BR(\chi^0_i\to \nu_{\ell} \tilde{\nu}_{\ell})} =
   \frac{(\tan\theta_W N_{i1}+N_{i2})^2}{(\tan\theta_W N_{i1}-N_{i2})^2}
   \left(\frac{m_{\chi^0_i}^2-m_{\tilde\ell}^2}
              {m_{\chi^0_i}^2-m_{\tilde{\nu}_{\ell}}^2} \right)^2 \; ,
\eeq                                                          
where $i$ is the label of the neutralino produced in the $\tilde c$ decay, 
and $N_{ij}$ are the elements of the unitary
matrix that diagonalizes the neutralino mass matrix in the $SU(2)\otimes U(1)$
gaugino basis~\cite{MSSM}. The dependence of this ratio on the MSSM parameters
will be studied in the following.
The decays of second and third generation sneutrinos are dominated by the
nearby pole of the virtual $\tilde{\nu}_e$, 
$\tilde\nu_i \to \nu_i \nu_e\tilde{\nu}_e$, and
give rise to a signature kinematically similar to that of the direct 
$\chi^0\to \nu_e \tilde{\nu}_e$ mode.
                 
In the case of $\tilde t$ production, the most promising decay mode for a CC
signal is driven by the chain:    
\beq                    
\tilde t \to b \chi^\pm \to 
        b \nu_{\ell} \tilde{\ell}_L \; ,
\eeq
followed by one of the possible decays of the charged slepton into
$\tilde{\nu}_e$. There is a competing (NC) decay of the $\chi^{\pm}$:
$\chi^{\pm}\to \ell \tilde{\nu}_{\ell}$. The relative branching ratios of the
two modes are given by:                                       
\beq
   \frac{BR(\chi^{\pm}_i\to \ell \tilde{\nu}_{\ell})}
        {BR(\chi^{\pm}_i\to \nu_{\ell} \tilde{\ell}_L)} =
   \frac{\vert U_{i1}\vert^2}{\vert V_{i1}\vert^2}
   \left(\frac{m_{\chi^{\pm}_i}^2-m_{\tilde\ell}^2}
              {m_{\chi^{\pm}_i}^2-m_{\tilde{\nu}_\ell}^2} \right)^2 \; ,
\eeq                                                          
where $i$ is the label of the chargino produced in the $\tilde t$ decay, 
and $U_{ij}$, $V_{ij}$ are the elements of the                  
matrices that diagonalize the chargino mass matrix~\cite{MSSM}.
For $\tan\beta=1$ the two decay rates are equal. For $\tan\beta>1$ the
$e\tilde\nu$ mode is always favoured. As we shall see in the following,
indenpedent reasons will select the region of $\tan\beta \sim 1$ as the most
favourable one.
\begin{figure}                                               
\centerline{\epsfig{figure=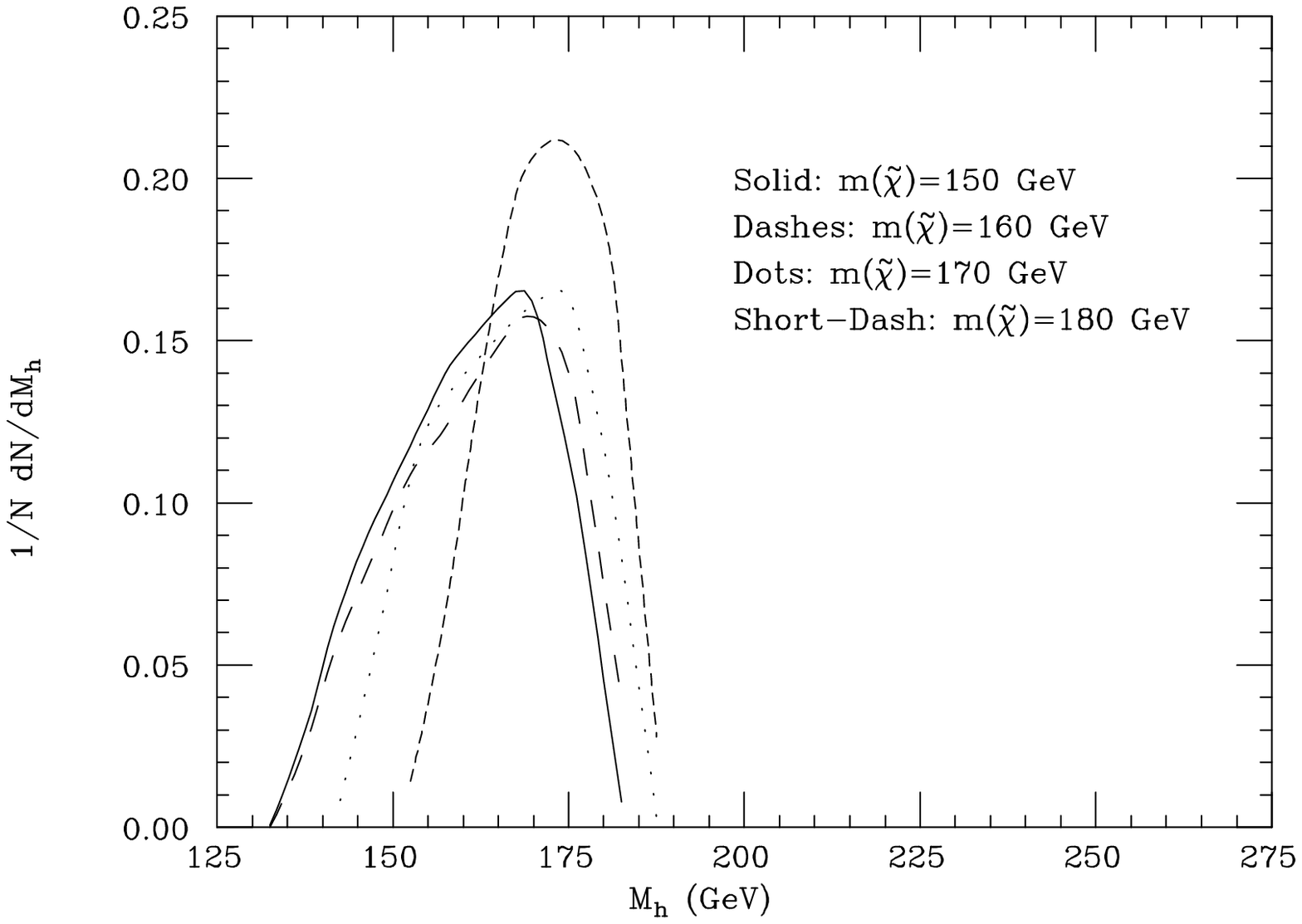,width=0.75\textwidth,clip=}}
\ccaption{}{ \label{fig:Mhcc} Distribution of the mass variable
$M_h$~\cite{h1,zeus,Blondel} for the CC
decay $\tilde c \to c \nu q \bar q$, for $m_{\tilde \nu}=60$~GeV and
different values of $m_{\chi^0}$.}
\end{figure}        
In these conditions, it is possible to verify that the dominant decay mode of
the $\tilde{\ell}_L$ is the one mediated by a virtual $W$, and leads
to $\tilde{\ell}_L \to \tilde{\nu}_{\ell} 
f\bar f^\prime$,  with $f,f^{\prime}$ an $SU(2)$ fermion doublet. The close
mass degeneracy between $\tilde{\ell}_L$ and $\tilde{\nu}$, due to low
$\tan\beta$, makes the overall transition $\tilde{\ell}_L\to \tilde{\nu}_e+X$
almost a ``1-body'' decay, with the unstable $\tilde{\nu}_e$ carrying away most
of the $\tilde{\ell}$ energy. The overall topology of the final state will
therefore appear very close to that of the $\tilde c \to c \nu \tilde{\nu}$
decay. For this reason we shall concentrate in the following on the kinematical
properties of the $\tilde c$ decays, since the CC decays of the stop
share the same overall features.

\subsection{Constraints on the $\tilde c$ case}
We now move on to discuss the constraints on the supersymmetric particle
spectrum dictated by the kinematical features of the possible
CC candidates reported by H1~\cite{h1}. 
The first constraints come from the recostructed mass spectrum of the four
CC high-$Q^2$ candidates. The presence of massive intermediate states in the
decay of the scalar quark significantly softens and smears out the mass of
the resonance, which can be reconstructed using the Jacquet--Blondel
variable $M_h$~\cite{Blondel}:
\beq          
   M_h=\sqrt{\frac{Q_h^2}{y_h}  }\; , \quad
   Q^2_h = \frac{P^2_{T,h}}{1-y_h}   \; , \quad
   y_h = \frac{\sum(E-P_z)}{2E_e^0} \;,
\eeq
where $E_e^0$ is the energy of the positron beam, and the sum extends over all
detectable energy in the final state.
For the sake of definiteness, we shall now concentrate
on the case of production and decay of the $\tilde c$. In the $\tilde c$ case, 
the  $M_h$~\cite{h1,zeus,Blondel}
distribution of the CC final states depends on the value of the neutralino and
of the sneutrino masses.  We show the dependence on the neutralino mass in
fig.~\ref{fig:Mhcc}, where we fixed $m_{\tilde \nu}=60$~GeV. The distributions,
normalized to unit area, were obtained by applying the H1 selection cuts 
$\missEt>50$~GeV and $Q^2>15000$ GeV$^2$. Higher values of the sneutrino mass
will soften each individual distribution. Lower values of the sneutrino mass
are excluded by the current LEP2 data, as will be shown in
the following.                                          
                                      
\begin{figure}                                               
\centerline{\epsfig{figure=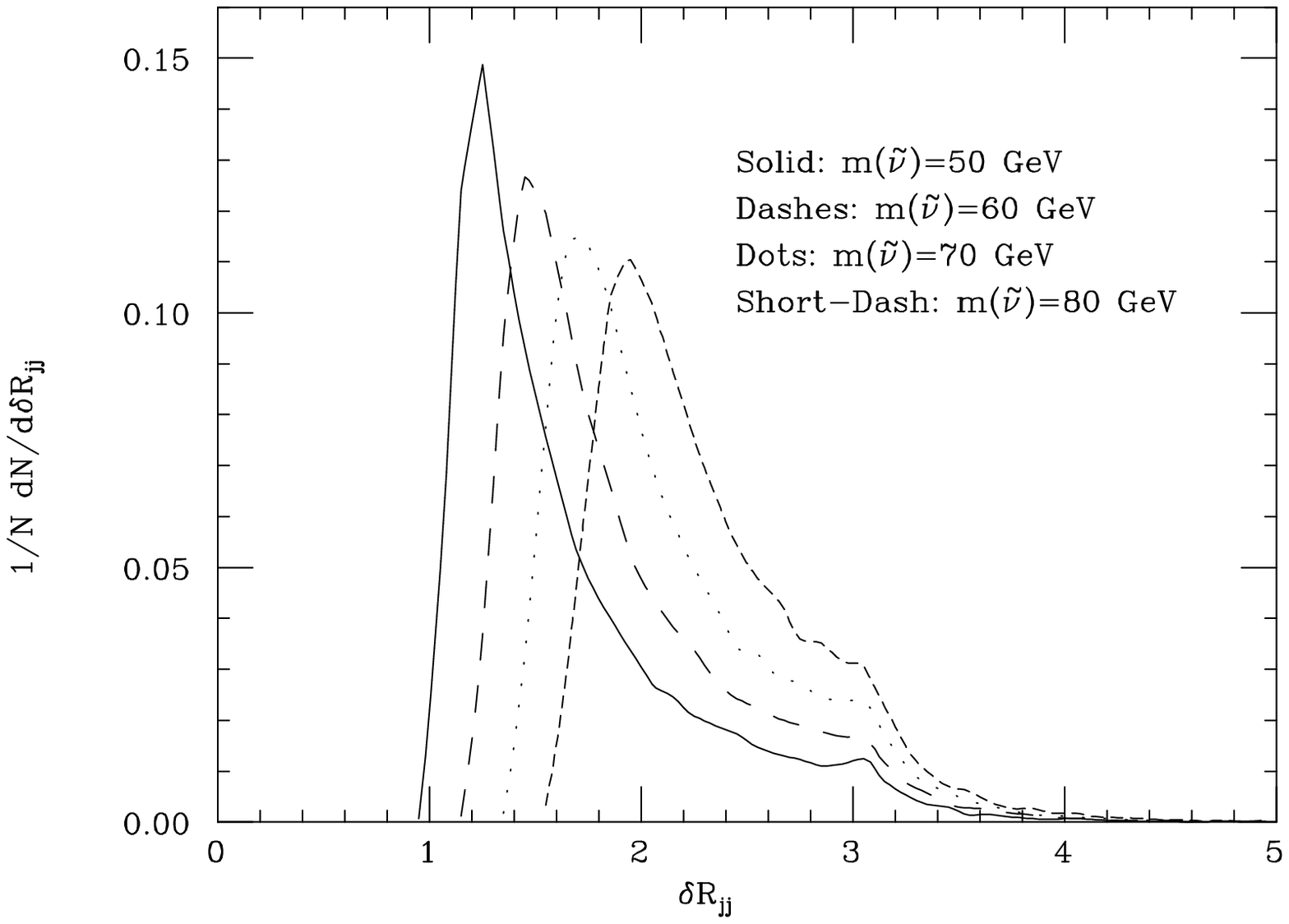,width=0.7\textwidth,clip=}}
\ccaption{}{ \label{fig:deltaR} Spatial separation of the two leading quarks
from the $\tilde c \to c \nu q\bar q$ decay, for $m_{\chi^0}=180$~GeV and
with various sneutrino masses.}
\end{figure}        
\begin{figure}                                               
\centerline{\epsfig{figure=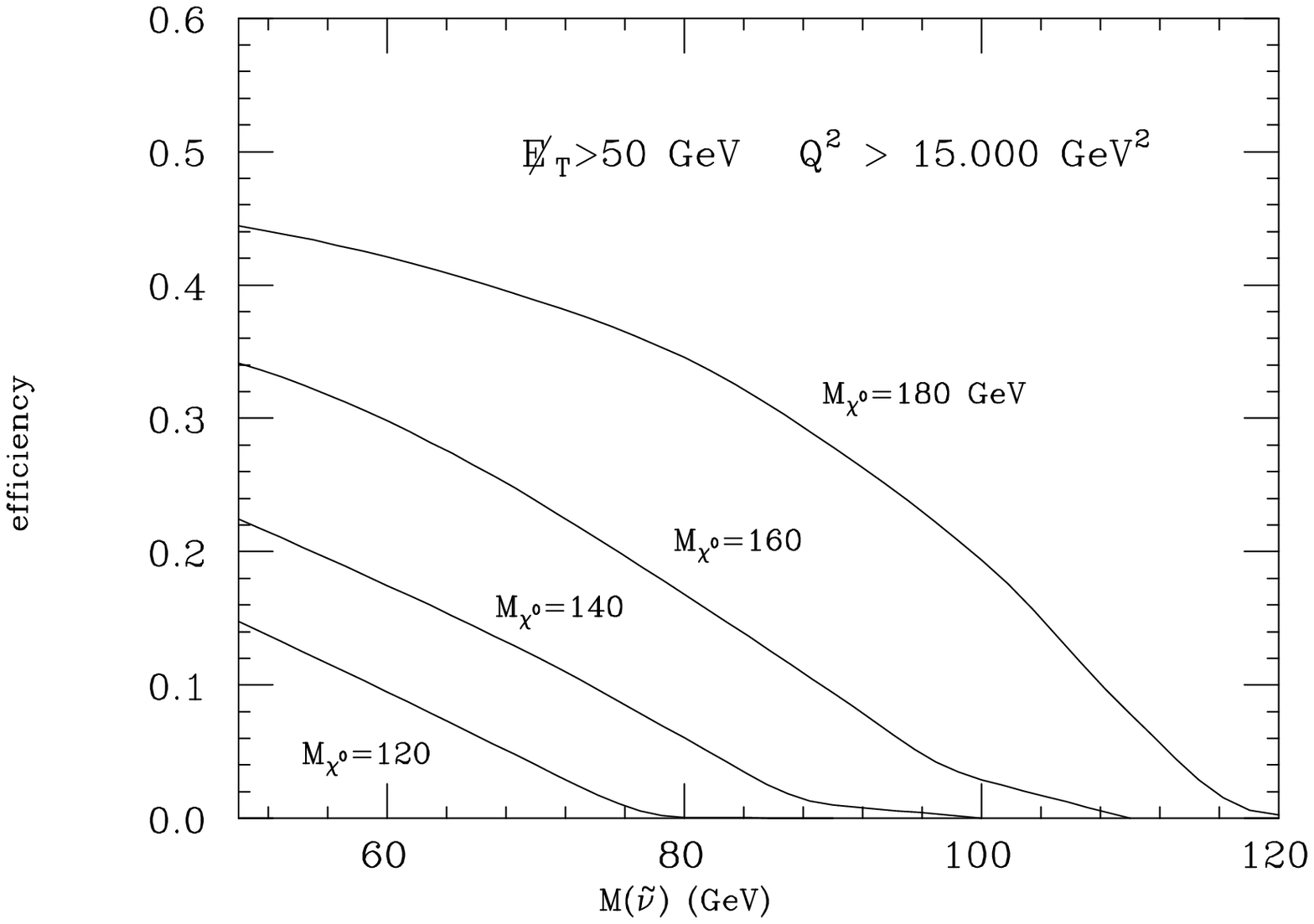,width=0.7\textwidth,clip=}}
\ccaption{}{ \label{fig:cceff} Charged current decay efficiencies.}
\end{figure}        

Comparing these distributions with the H1 data (see table~7 of ref.~\cite{h1}),
and taking into account the quoted 10\% relative uncertainty on the measured  
$M_h$, we conclude that only neutralino masses in excess of 170--180~GeV have a
chance of producing the observed signals.

Additional information of the spectrum comes from the study of the topology of
the final-state hadronic system. 
With such a large neutralino mass, the charm jet produced in the $\tilde c$
decay would be very soft and broad. The most energetic hadronic activity in the
event would come from the $R$-parity violating two-quark decay of the 
sneutrino. 
A light sneutrino could result in the two jets being merged into a
single broad one, while a heavy sneutrino would result in two widely 
separated jets. These alternatives are illustrated in fig.~\ref{fig:deltaR},
where we plot the distribution of the $\eta-\phi$ separation ($\delta R$)
between the two  highest-$E_T$ partons in the events. As the figure shows,
experimental evidence for a hadronic final state mostly consisting of a single
broad jet would require 
a sneutrino mass as light as possible and not exceeding 70~GeV.

Similar conclusions can be reached by a study of the event rate. The fraction
of events which satisfies the H1 selection cuts
is plotted in fig.~\ref{fig:cceff}, as a function of the sneutrino mass 
and for different choices of the neutralino mass. 
Once again, acceptable efficiencies can be obtained only assuming a large
neutralino mass and a small sneutrino mass.

\begin{figure}                                               
\centerline{\epsfig{figure=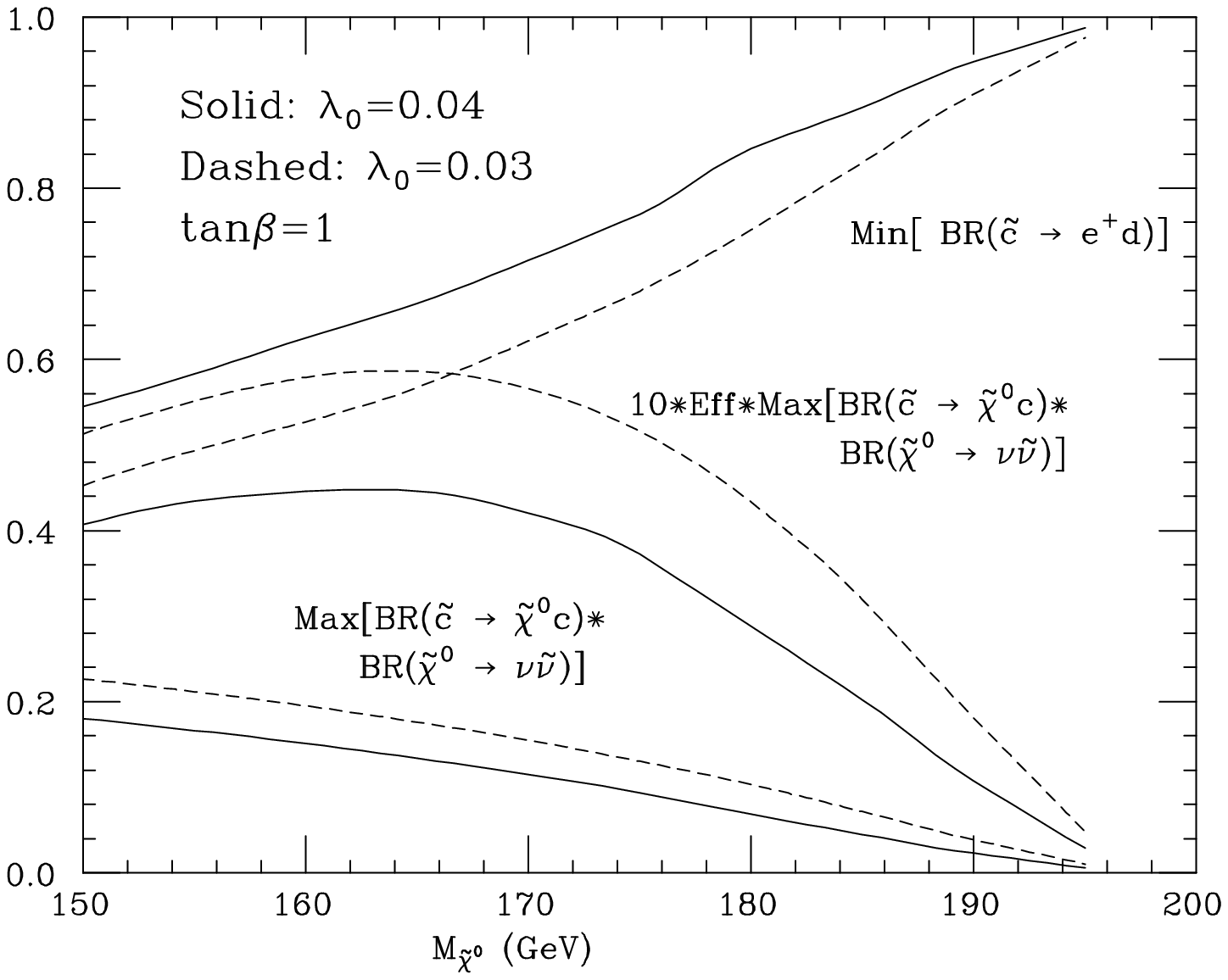,width=0.75\textwidth,clip=}}
\ccaption{}{ \label{fig:CCBr} Decay branching ratios of relevance to the CC
$\tilde c$ decay, as a function of the neutralino mass. The uppermost curves
correspond to the lowest possible value of \BReq\ consistent with
$m_{\chi^\pm}>200$~GeV and $\tan\beta=1$.  The lowest set of curves gives the
maximum combined branching ratio for the CC $\tilde c$ decay, under the same
assumptions. The central set of curves gives the product of the kinematic
efficiency (shown in fig.~\ref{fig:cceff}) times the overal CC branching ratio,
rescaled by a factor of 10.}
\end{figure}               

We now proceed to verifying whether these kinematical constraints allow for
acceptable branching ratios into the desired CC final states.
Figure~\ref{fig:CCBr} shows, as a function of the neutralino mass and for
$\tan\beta=1$, the minimum
value of the branching ratio for the NC decay $\tilde c \to e^+ d$ allowed by
scanning the $(\mu,M_2)$ plane under the assumption of gaugino
mass unification and that the mass of the
lightest chargino be larger than 200~GeV. We present two curves,
corresponding to the choices of the coupling 
$      \lambda^{\prime}\sqrt{{\cal B}_{ed}}
      \equiv \lambda^{d}_0= 0.03$ and 0.04.
The constraint ${\cal B}_{ed}<0.75$
excludes neutralino masses larger than approximately 180~GeV. 
As a result we see that the neutralino mass value is
fixed to be in the range 170--180~GeV, the lower limit being dictated by the
kinematics, and the upper limit being defined by the branching ratio
requirements. 
The APV constraint $\BReq>0.5$ is automatically satisfied in the
neutralino mass range $m_{\chi^0}>170$~GeV, as fig.~\ref{fig:CCBr} shows.

The previous results were obtained using $\tan\beta=1$. 
Similar results can be obtained increasing the value of $\tan\beta$. The
maximization of the rate would select neutralinos with a higher bino content.

\begin{figure}                                               
\centerline{\epsfig{figure=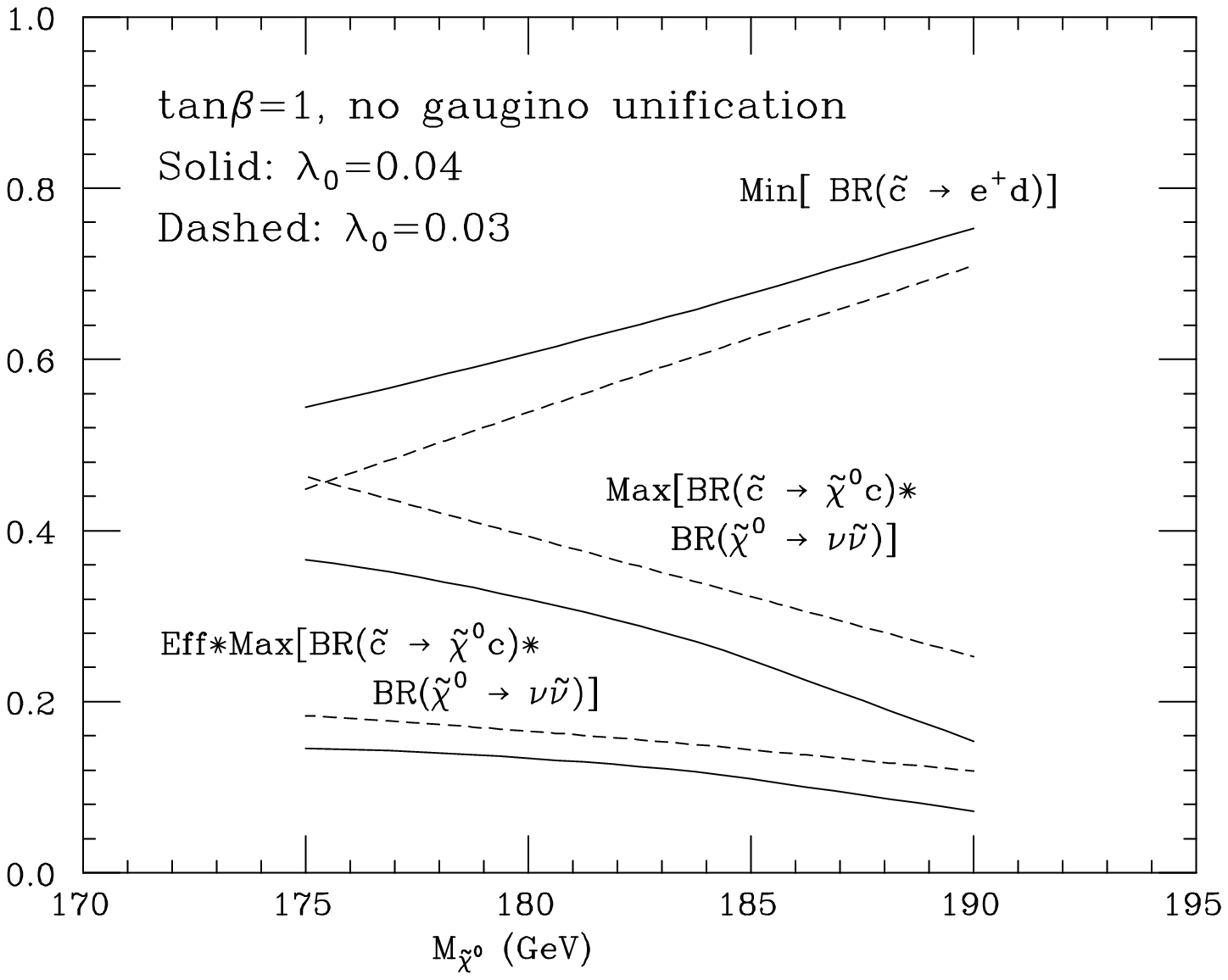,width=0.75\textwidth,clip=}}
\ccaption{}{ \label{fig:CCBrm1} Decay branching ratios of relevance to the CC
$\tilde c$ decay, as a function of the neutralino mass, in absence of
the gaugino-mass
unification hypothesis. The upper curves
correspond to the lowest possible value of \BReq\ consistent with
$m_{\chi^\pm}>200$~GeV and $\tan\beta=1$.  The central set of curves gives the
maximum combined branching ratio for the CC $\tilde c$ decay, under the same
assumptions. The lowest set of curves gives the product of the kinematic
efficiency (shown in fig.~\ref{fig:cceff}) times the overal CC branching
ratio.}
\end{figure}               

If we take into account the possible alternative neutralino decay mode,
$\chi^0\to \ell \tilde \ell$, and assume the mass degeneracy of $\tilde\ell_L$
and $\tilde\nu_L$ resulting from $\tan\beta=1$ and $SU(2)\otimes U(1)$
invariance, a scan of the $(\mu,M_2)$ plane leads to the upper value of the
combined branching ratios ${\cal B}(\tilde c \to c \chi^0)\times 
{\cal B}(\chi^0 \to \nu \tilde \nu)$ given in fig.~\ref{fig:CCBr}.
The maximum attainable values for such branching ratio are around 15\% in the
relevant neutralino mass range. Combined with the efficiency of the selection
cuts, this results in an overall fraction of detectable CC events of about
5\%, assuming a sneutrino of 60~GeV, namely as light as currently allowed by
LEP2. This should be compared to a combined efficiency (64\%) 
times branching ratio (70\%)                                  
for the NC $\tilde c$ final states of about 45\%. 
The predicted relative rate of
CC over NC events is therefore of the order of 1/9. This low ratio would
predict only a fraction of a CC event in the current data sample,
and it would require a significant                                  
statistical fluctuation in the observed rates for this scenario to be tenable.

A more optimistic conclusion could be reached by relaxing some of the
underlying MSSM assumptions. For example, one could remove the constraint of
gaugino-mass unification. The situation in this case is shown in
fig.~\ref{fig:CCBrm1}, where we plot the maximum possible CC branching ratio 
obtained by scanning the parameters $(\mu, M_1, M_2)$ with the
$m_{\chi^\pm}>200$~GeV constraint. The optimal condition is reached when
the gaugino mass parameters $M_1$ and $M_2$ are comparable in size.
Notice that in this case on can benefit from both 
a larger $BR(\tilde c\to c\chi^0)$, and from a 
$BR(\chi^0\to \nu\tilde{\nu})$ close to 1. The relative rate CC/NC can attain
values as large as 1/3, consistent with the reported CC excess. We point out
that in this case the interesting signature of same-sign dileptons at the
Tevatron, typical of conventional R-conserving decays,
would be significantly suppressed.                              

\begin{figure}                                               
\centerline{\epsfig{figure=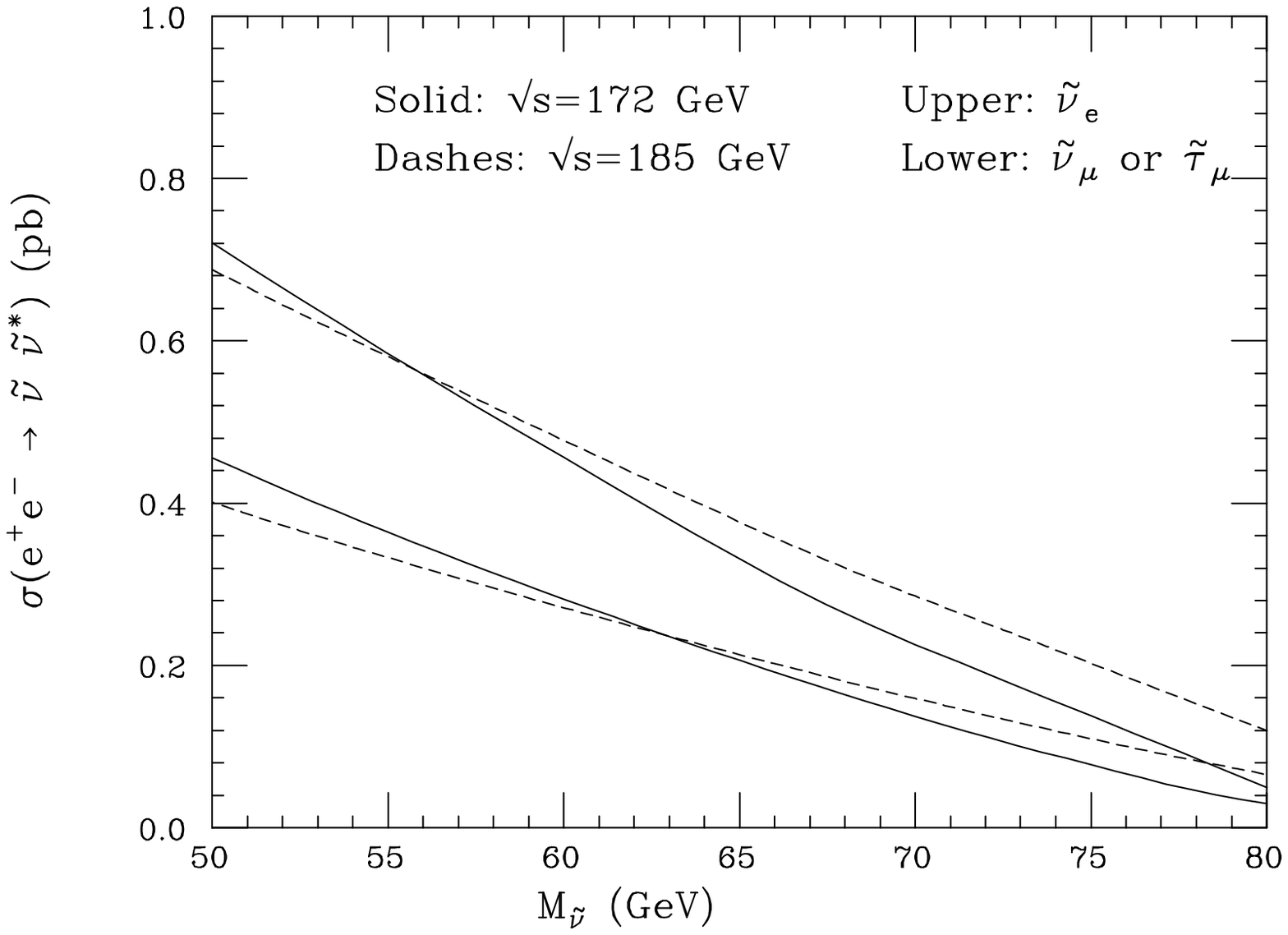,width=0.7\textwidth,clip=}}
\ccaption{}{ \label{fig:snuxs} Sneutrino cross section at $\sqrt{s}=172$ and
185~GeV. The $\tilde\nu_{\mu}$ cross section is independent of supersymmetric
parameters,
and is clearly equal to the $\tilde{\nu}_{\tau}$ cross section.
For the $\tilde{\nu}_e$ cross section we chose $\tan\beta=1$ and 
the $(\mu,M_2)$ values which
maximise the overall $\tilde c$ CC decay rate. ISR corrections are
included~\cite{Mangano96}.}
\end{figure}

We point out once more that for this scenario to have any chance of working, a
sneutrino mass just above the current LEP2 constraints is required, and its
discovery at the coming high-statistics $\sqrt{s}=185$~GeV run of LEP2 should
be granted. For illustration, we show in fig.~\ref{fig:snuxs} the sneutrino
$e^+e^-$ production cross section at $\sqrt{s}=172$ and 185~GeV, obtained by
assuming the costraints on the values of $\mu$ and $M_2$ set by the previous
analyses. For reference, the current cross section limit on 4-jet final states
from the decay of a pair of resonances in the mass range 50--60~GeV is about
0.5~pb~\cite{Schlatter}.                                               
Notice however that Aleph~\cite{Aleph}
sees a 4-jet signal above background, which
in principle could be explained by 
pair production of $\tilde{\nu}_e$ followed by $\tilde{\nu}\to jj$.
                                                                 
\begin{figure}                                               
\centerline{\epsfig{figure=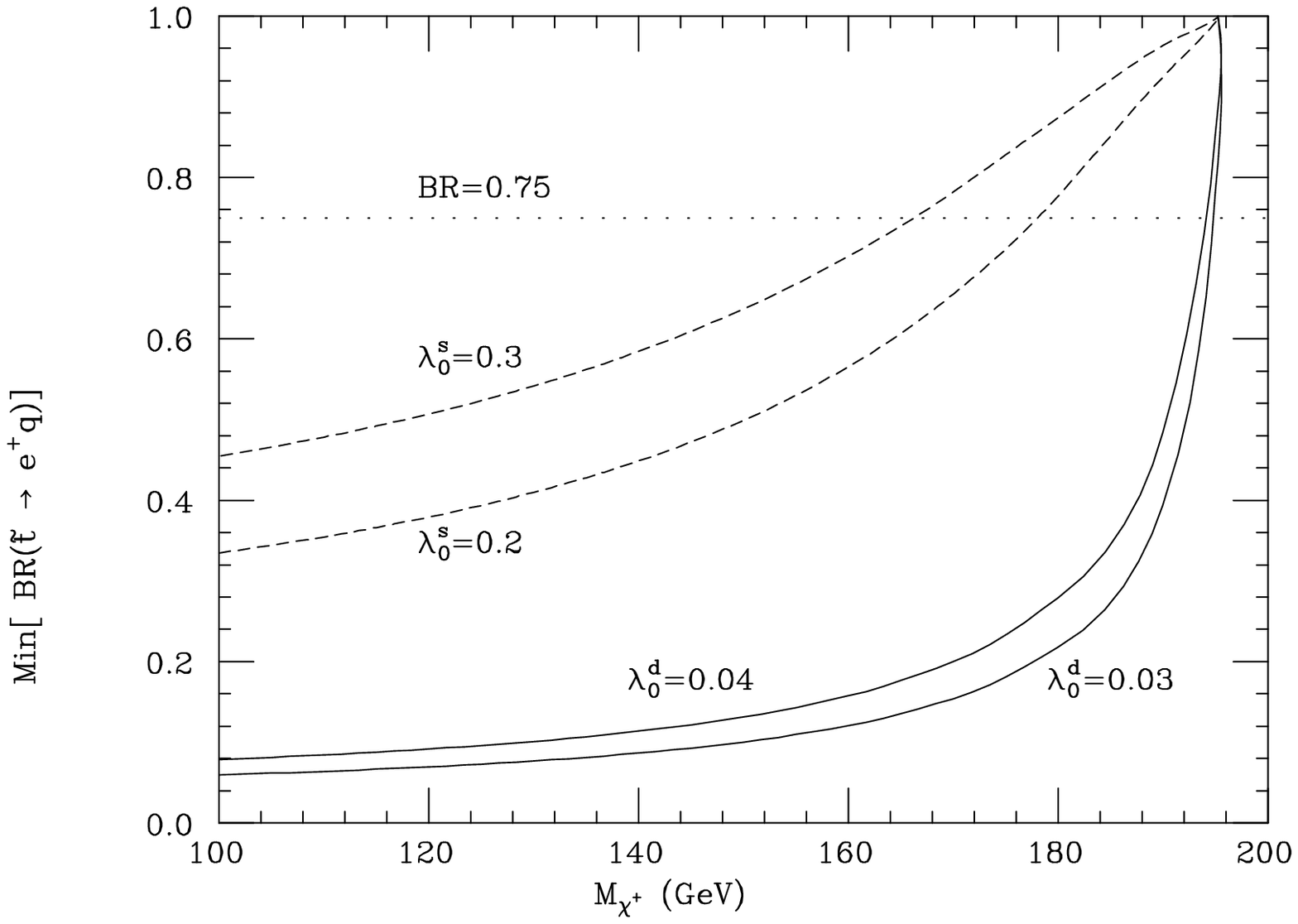,width=0.7\textwidth,clip=}}
\ccaption{}{ \label{fig:stopbr} Minimum $\tilde t \to e^+ q$ branching ratios,
as a function of the chargino mass and of the R-violating coupling strength
$\lambda_0^{q}$, for $\tan\beta=1$. The branching ratios are minimised 
for a gaugino-like chargino.}
\end{figure}                          

\subsection{Constraints on the $\tilde t$ case}
The kinematical constraints on the final states of $\tilde t$ production are
analogous to those studied in the case of $\tilde c$ production and decay. In
particular, even in this case it is important that the chargino mass be as
large as possible, compatibly with an acceptable value of \BReq. 
The value of \BReq\ for $\tilde t$ decays is plotted in fig.~\ref{fig:stopbr}
as a function of the chargino mass. 
The kinematical
constraint on the allowed range of gaugino masses is consistent with the
requirements set by the branching ratios in  the case of the
$R$-violating coupling of the $\tilde t$ to $e^+d$. In particular, 
chargino masses in the range of $180-190$~GeV are acceptable. The
inclusion of mixing in the $\tilde t_L-\tilde t_R$ system, which leaves
unaltered both the kinematics and the combined Tevatron and APV constraints on
\BReq, modifies this range only marginally. In the case of coupling to 
$e^+s$, the Tevatron limits exclude the reference value of $\lambda_0^{s}=0.3$,
and allow chargino masses up to 175~GeV for                                    
$\lambda_0^{s}=0.2$.
                  
As already discussed, the CC final state for $\tilde t$ decay is obtained via
the transition $\tilde t \to b \chi^{\pm}$, followed by $\chi^{\pm} \to
\nu\tilde \ell^{\pm}$. As already remarked, large efficiencies and an $M_h$
spectrum consistent with that of the potential H1 candidates require a slepton
as light as possible. This forces the selectron mass not to exceed significantly
the value of 60~GeV. Given the LEP2 limit on the sneutrino mass, which is also
at the level of 60~GeV, we conclude from eq.~(\ref{eq:massrel}) that $\tan\beta$
should be very close to 1.

A chargino mass of about 190~GeV allows, in the case of coupling to the $d$
quark, values of \BReq\ as low as 50\%, which is the limit
permitted by the APV constraints. Assuming a 50\% branching ratio for the
$\chi^+\to \nu \tilde{\ell}^+$ decay (consistent with the assumption of
$\tan\beta$ close to 1), and using a typical kinematical
efficiency of 50\%, we obtain an overall fraction of 12\% CC final
states passing the H1 cuts. This can be compared to the $64\%({\rm efficiency})
\times 50\%({\rm branching~ ratio})=32\%$                       
fraction of predicted NC events. The relative CC/NC rate can therefore be as
large as 1/3, which is well consistent with the H1 indications.

\section{Conclusions}
Even in absence of additional constraints from the CC channel, it is not easy
to incorporate the possible indications of new physics from HERA in our present
theoretical understanding. For example, contact terms require values of
$g^2/\Lambda^2 \sim 4\pi/(3~{\rm TeV})^2$, which would imply a very strong
nearby
interaction. Indeed for $g^2$ of the order of the $SU(3)\bigotimes SU(2)
\bigotimes U(1)$ couplings, $\Lambda$ would fall below 1~TeV, where the contact
term description is inadequate. Squarks with $R$-parity violation are perhaps
the most appealing version of leptoquarks. However they require a very peculiar
family and flavour structure, whose pattern can
be embedded~\cite{Barbieri97} in a grand       
unification framework.
The already intricated problem of the
mysterious texture of masses and couplings is however
terribly enhanced in these scenarios.
                                     
If CC events are observed in the same range of $Q^2$ at HERA at a roughly
comparable rate, then most of the models so far considered for the
interpretation of NC events are to be reconsidered. For example, we could not
find an acceptable set of contact terms compatible with $SU(2)\bigotimes U(1)$ 
invariance.
In the case of leptoquarks, only peculiar models
with $SU(2)\bigotimes U(1)$ breaking couplings
or with charged currents involving the $c$ quark survive. The CC events show in
this case only one-parton jets in the final state and the branching ratio for
CC decays could well be of the order of 50\%. 

Squarks could produce CC events
with multi-parton final states. Clean one-jet final states, as observed in the 
NC events,                                         
would therefore require neutralinos (in the $\tilde c$ case) or charginos  
(in the $\tilde t$ case) with a mass very close to 200~GeV, and sleptons as
light as possible. 
The additional constraints set by the Tevatron and APV limits 
on the branching ratios limit the neutralino mass to the range $170-180$~GeV.
If the gaugino-unification relation holds, at best 
one can hope for the relative rate of
CC to NC events to be around 1/9.  Relaxing the gaugino-mass unification
hypothesis allows this ratio to increase to the acceptable value of 1/3.
                                                                      
Similar kinematic constraints exist in the case of the $\tilde t$. In this
case, one is also forced to assume the $\tilde e_L$ mass to be as light as
possible, forcing the value of $\tan\beta$ to be around 1.
Acceptable values of the chargino mass are in the range of
$180-190$~GeV, if one assumes the $e^+d\to \tilde t$ production mechanism. The
$e^+ s \to \tilde t$ case leads instead to lower values of the chargino mass,
which are not obviously consistent with the $M_h$ spectrum of the H1 CC
candidate events. The CC/NC ratio can be as large as 1/3.
In all cases, one is left with the prediction that the sneutrino (and possibly
the left selectron as well) should be within the reach of the coming LEP2 runs.

We look
forward to the results of the ongoing run of HERA, that will hopefully clarify
to some extent the present situation.                                  
\\[0.5cm]
\noindent
{\bf Acknowledgements.} We would like to thank J. Ellis for useful discussions
and comments.
                                                            
\def\ijmp#1#2#3{{\it Int. Jour. Mod. Phys. }{\bf #1~}(19#2)~#3}
\def\pl#1#2#3{{\it Phys. Lett. }{\bf B#1~}(19#2)~#3}
\def\zp#1#2#3{{\it Z. Phys. }{\bf C#1~}(19#2)~#3}
\def\prl#1#2#3{{\it Phys. Rev. Lett. }{\bf #1~}(19#2)~#3}
\def\rmp#1#2#3{{\it Rev. Mod. Phys. }{\bf #1~}(19#2)~#3}
\def\prep#1#2#3{{\it Phys. Rep. }{\bf #1~}(19#2)~#3}
\def\pr#1#2#3{{\it Phys. Rev. }{\bf D#1~}(19#2)~#3}
\def\np#1#2#3{{\it Nucl. Phys. }{\bf B#1~}(19#2)~#3}
\def\mpl#1#2#3{{\it Mod. Phys. Lett. }{\bf #1~}(19#2)~#3}
\def\arnps#1#2#3{{\it Annu. Rev. Nucl. Part. Sci. }{\bf #1~}(19#2)~#3}
\def\sjnp#1#2#3{{\it Sov. J. Nucl. Phys. }{\bf #1~}(19#2)~#3}
\def\jetp#1#2#3{{\it JETP Lett. }{\bf #1~}(19#2)~#3}
\def\app#1#2#3{{\it Acta Phys. Polon. }{\bf #1~}(19#2)~#3}
\def\rnc#1#2#3{{\it Riv. Nuovo Cim. }{\bf #1~}(19#2)~#3}
\def\ap#1#2#3{{\it Ann. Phys. }{\bf #1~}(19#2)~#3}
\def\ptp#1#2#3{{\it Prog. Theor. Phys. }{\bf #1~}(19#2)~#3}
\def\ZPC#1#2#3{{\sl Z.~Phys.} {\bf C#1}~(#3) #2}
\def\PTP#1#2#3{{\sl Prog. Theor. Phys.} {\bf #1}~(#3) #2}
\def\PRL#1#2#3{{\sl Phys. Rev. Lett.} {\bf #1}~(#3) #2}
\def\PRD#1#2#3{{\sl Phys. Rev.} {\bf D#1}~(#3) #2}
\def\PLB#1#2#3{{\sl Phys. Lett.} {\bf B#1}~(#3) #2}
\def\PREP#1#2#3{{\sl Phys. Rep.} {\bf #1}~(#3) #2}
\def\NPB#1#2#3{{\sl Nucl. Phys.} {\bf B#1}~(#3) #2}
                        

\begin{thebibliography}{99}
          
\bibitem{h1}
  C. Adloff {\it et al.}, H1 collaboration, DESY 97-24, 
  {\tt hep-ex/9702012}.
\bibitem{zeus}        
  J. Breitweg {\it et al.}, ZEUS collaboration, DESY 97-25, 
  {\tt hep-ex/9702015}.                              
\bibitem{Altarelli97}
  G. Altarelli, J. Ellis, G.F. Giudice, S. Lola, and M.L. Mangano, CERN
  TH/97-40, {\tt hep-ph/9703276}.
\bibitem{HERARPV}
  D. Choudhury and S. Raychaudhuri, CERN TH/97-26, {\tt hep-ph/9702392};\\
  H. Dreiner and P. Morawitz,  {\tt hep-ph/9703279};\\
  J. Kalinowski, R. Ruckl, H. Spiesberger, and P.M. Zerwas, 
  BI-TP-97-07, {\tt hep-ph/9703288};\\                      
  T. Kon and T. Kobayashi, ITP-SU-97-02, {\tt hep-ph/9704221}.
\bibitem{HERAlq}                                              
  J.L. Hewett and T.G. Rizzo, SLAC-PUB-7430, {\tt hep-ph/9703337}.      
\bibitem{HERAcomp1}
  K.S. Babu, C. Kolda, J. March-Russell, and F. Wilczek, 
  IASSNS-HEP-97-04, {\tt hep-ph/9703299}.
\bibitem{HERAcomp2}                      
  V. Barger, K. Cheung, K. Hagiwara and D. Zeppenfeld, MADPH-97-991, {\tt
  hep-ph/9703311};\\                      
  N. Di Bartolomeo and M. Fabbrichesi, SISSA-34-97-EP, {\tt hep-ph/9703375}.
\bibitem{HERAcomp3}
  M.C. Gonzalez-Garcia and S.F. Novaes, IFT-P-024-97, {\tt hep-ph/9703346};\\
  A.E. Nelson, {\tt hep-ph/970337};\\
  W. Buchm\"uller and D. Wyler, DESY-97-066, {\tt hep-ph/9704317}.
\bibitem{Opal96}
 G. Alexander {\it et al.}, OPAL collaboration, CERN-PPE/96-156.
\bibitem{Opal97}
 S. Komamiya, for the OPAL collaboration, CERN seminar, Feb. 25th, 1997,
\\
{\tt http://www.cern.ch/Opal/plots/komamiya/koma.html}.
\bibitem{Bodek96}                                      
  A. Bodek, for the CDF collaboration, FERMILAB-Conf-96/381-E, Proc.
  of Cracow International Symposium on Radiative Corrections, Poland, 1996.
\bibitem{wood} C.S. Wood {\it et al.}, {\it Science} {\bf 275} (1997) 1759.
\bibitem{mas} 
W.J. Marciano and J.L. Rosner, \prl{65}{90}{2963}, Erratum \prl{68}{92}{898};\\
B.P. Masterson and C.E. Wieman, in {\it Precision Tests
of the Standard Electroweak Model}, ed. P.~Langacker (World Scientific,
Singapore,1995).
\bibitem{pdg} Particle Data Group, R.M. Barnett {\it et al.}, 
\pr{54}{96}{1}.
\bibitem{mar} W.J. Marciano and A. Sirlin, \prl{71}{93}{3629};\\
M. Finkemeier, \pl{387}{96}{391}.
\bibitem{sha} O. Shanker, \np{204}{82}{375}.
\bibitem{vol} M.B. Voloshin, \pl{283}{92}{120}.
\bibitem{Bluemlein96}
  J. Bl\"umlein, E. Boos and A. Kryukov,
  {\tt hep-ph/9610408};\\
  J. Bl\"umlein, {\tt hep-ph/9703287}. 
\bibitem{D0LQ}                                           
 D0 collaboration, 
{\tt http://d0wop.fnal.gov/public/new/lq/lq\_blurb.html}.
\bibitem{abr} H. Abramowicz {\it et el.}, \zp{12}{82}{225}.
\bibitem{david} S. Davidson, D. Bailey and B.A. Campbell, \zp{61}{93}{613}.
\bibitem{MSSM}
 For reviews, see, for instance: \\ 
 H.P. Nilles, \prep{110}{84}{1}; \\
 H.E. Haber and G.L. Kane, \prep{117}{85}{75}; \\
 G.G. Ross, {\it Grand Unified Theories}, (Benjamin-Cummings, Menlo Park,
 CA, 1985); \\
 R. Barbieri, Riv. Nuovo Cimento {\bf 11}\rm   (1988) 1.
\bibitem{RPth}  G. Farrar and P. Fayet, \pl{76}{78}{575}; \\
 S. Weinberg, \pr{26}{82}{287}; \\                    
 N. Sakai and T. Yanagida, \np{197}{82}{133}.
\bibitem{Hewett}
J.L. Hewett, Proc. 1990 Summer Study on High Energy Physics, Snowmass,
Colorado.
\bibitem{Kon}
T. Kon and T. Kobayashi, \PLB{270}{81}{1991};\\
T. Kon, T. Kobayashi, S. Kitamura, K. Nakamura and S. Adachi,
\ZPC{61}{239}{1994};\\
T. Kobayashi, S. Kitamura, T. Kon,
{\it Int. J. Mod. Phys.} {\bf A11} (1996) 1875.                         
\bibitem{Dreiner}
J. Butterworth and H. Dreiner, \NPB{397}{3}{1993}; \\
H. Dreiner and P. Morawitz, \NPB{428}{31}{1994}; \\
E. Perez, Y. Sirois and H. Dreiner, contribution to Beyond the Standard
Model Group, 1995-1996 Workshop on Future Physics at HERA, see also the
Summary by \\
H. Dreiner, H.U. Martyn, S. Ritz and D. Wyler, {\tt hep-ph/9610232}.
\bibitem{Kunszt97}
  Z. Kunszt and W.J. Stirling, DTP-97-16, {\tt hep-ph/9703427};\\
  T. Plehn, H. Spiesberger, M. Spira and P.M. Zerwas,              
  DESY-97-043, {\tt hep-ph/9703433}.
\bibitem{Spira96}                                                  
  W. Beenakker, R. Hopker, M. Spira and P.M. Zerwas, {\tt hep-ph/9610490}.
\bibitem{Spira97}
 M. Kramer, T. Plehn, M. Spira and P.M. Zerwas,
 RAL-97-017, {\tt hep-ph/9704322}.
\bibitem{Blondel}
  F. Jacquet and A. Blondel, in Proceedings of the Study for an $ep$ facility
  for Europe, U. Amaldi ed., DESY--79--48 (1979) p.391.
\bibitem{Mangano96}
M.L. Mangano, G. Ridolfi {\it et al.}, {\tt hep-ph/9602203}, published
in Physics at LEP~2, eds. G. Altarelli, T. Sj\"ostrand and F. Zwirner, CERN
Report 96-01, vol. 2, p.299.
\bibitem{Schlatter}
W.D. Schlatter, for the LEP Working Group on Four-Jet Final States, CERN
seminar, Feb. 25th, 1997.
\bibitem{Aleph}
D. Buskulic {\it et al.}, ALEPH collaboration, \ZPC{71}{179}{1996}, \\
G. Cowan, for the ALEPH collaboration,                         
CERN seminar, Feb. 25th, 1997 \\
{\tt http://alephwww.cern.ch/ALPUB/seminar/Cowan-172-jam/cowan.html}.
\bibitem{Barbieri97}
 R. Barbieri, A. Strumia and Z. Berezhiani, IFUP-TH-13-97, {\tt
 hep-ph/9704275};\\
 G.F. Giudice and R. Rattazzi, CERN-TH-97-076, {\tt hep-ph/9704339}.
\end{thebibliography}
\end{document}